# Discrimination of the DPRK underground explosions and their aftershocks using the P/S spectral amplitude ratio

Ivan O Kitov and Mikhail V. Rozhkov

**Abstract**

We have estimated the performance of discrimination criterion based on the P/S spectral amplitude ratios obtained from six underground tests conducted by the DPRK since October 2006 and six aftershocks induced by the last two explosions. Two aftershocks were detected in routine processing at the International Data Centre of the Comprehensive Nuclear-Test-Ban Treaty Organization. Three aftershocks were detected by a prototype waveform cross correlation procedure with explosions as master events, and one aftershock was found with the aftershocks as master event. Two seismic arrays USRK and KSRS of the International Monitoring System (IMS) and two non-IMS 3-C stations SEHB (South Korea) and MDJ (China) were used. With increasing frequency, all stations demonstrate approximately the same level of deviation between the $P_g/L_g$ spectral amplitude ratios belonging to the DPRK explosions and their aftershocks. For a single station, simple statistical estimates show that the probability of any of six aftershocks not to be a sample from the explosion population is larger than 99.996% at the KSRS and even larger at USRK. The probability of any of the DPRK explosion to be a representative of the aftershock population is extremely small as defined by the distance of 20 and more standard deviations to the mean explosion $P_g/L_g$ value. For network discrimination, we use the Mahalanobis distance combining the $P_g/L_g$ estimates at three stations: USRK, KSRS and MDJ. At frequencies above 4 Hz, the (squared) Mahalanobis distance, $D^2$, between the populations of explosions and aftershocks is larger than 100. In the frequency band between 6 and 12 Hz at USRK, the aftershocks distance from the average explosion $D^2 > 21,000$. Statistically, the probability to confuse explosions and aftershocks is negligible. These discrimination results are related only to the aftershocks of the DPRK tests and cannot be directly extrapolated to the population of tectonic earthquakes in the same area.

**Introduction**

The main difference between seismic sources of underground explosions and earthquakes, which is useful for discrimination, is related to the efficiency of P- and S-wave generation. Explosions demonstrate more efficient generation of compressional (P) waves, while earthquakes generate shear waves (S) of relatively higher amplitude. As a result, the difference between body (P) wave, **m**$_b$, and surface (related to S) wave, **M**$_s$, magnitudes measured at teleseismic distances successfully serves for discrimination of earthquakes and underground explosions.

At regional distances, seismic wave-field is characterized by a much higher fluctuation in amplitudes of P and S-waves than that at teleseismic distances because of inhomogeneous velocity and attenuation structures of the crust and the uppermost mantle. No reliable magnitude scale can be derived for discrimination purposes. Therefore, finer characteristics are needed to distinguish between earthquakes and explosions. It was found that the spectrum of P-wave falls faster for earthquakes than that of explosion sources. Based on this observation, a spectral discriminant was introduced as a frequency dependent ratio of P- and S-wave spectral amplitudes. In the past, we conducted a thorough study of spectral discriminants and other discrimination methods, such as statistical identification, artificial neural networks classification and moment tensor estimates, using data from the Soviet nuclear tests, regional chemical explosions and earthquakes [1-7]. It was tested in a number of geological/seismological regions (chiefly around test sites and PNEs) and demonstrated good performance at the level of (teleseismic) $M_s$-$m_b$ criterion, but for much smaller events, which are not measured at teleseismic stations.

**Data and method**

We test the P/S spectral amplitude ratio criterion as a possible tool to discriminate six underground tests conducted by the DPRK since October 2006 (see Table 1) and six smaller events found near these explosions just after the DPRK5 and DPRK6. These six smaller events were detected by waveform cross correlation at regional stations of the International monitoring system (IMS) of the Comprehensive Nuclear-Test-Ban Treaty Organization (CTBTO) and non-IMS stations (see Table 2). Since their waveforms demonstrate high correlation coefficients with those from the DPRK explosions and the arrival times found by cross correlation do not depend on the length of correlation window one may consider these low-magnitude events as located very close or within the DPRK tests site. Taking into account that the smaller events occurred just after the two biggest tests, their low magnitudes (2 and more units of magnitude lower than their corresponding main shocks), and the spatial closeness with the tests, one cannot reject the hypothesis that these are aftershocks of the DPRK5 and DPRK6.

Table 1. Parameters of six DPRK tests estimated by the IDC

| Event | Lat, deg | Lon, deg | ndef | $m_b$ (IDC) | $M_S$ (IDC) | $M_L$ (IDC) |
|---|---|---|---|---|---|---|
| DPRK1 | 41.312 | 129.02 | 22 | 4.08 | - | 3.89 |
| DPRK2 | 41.311 | 129.05 | 72 | 4.51 | 3.56 | 4.27 |
| DPRK3 | 41.301 | 129.07 | 110 | 4.92 | 3.95 | 4.52 |
| DPRK4 | 41.304 | 129.05 | 102 | 4.82 | 3.92 | 4.61 |
| DPRK5 | 41.299 | 129.05 | 120 | 5.09 | 4.17 | 4.29 |
| DPRK6 | 41.32 | 129.03 | 189 | 6.07 | 4.91 | 5.17 |

ndef - the number of defining phases

The IMS seismic network detected six underground tests conducted by the DPRK between October 9, 2006 and September 3, 2017. The number of seismic stations detecting signals from the DPRK tests and the number of phases associated with the explosions, *ndef*, has been gradually increasing with the growth in the number of working IMS seismic stations as well as in the test magnitude. As a result of the event magnitude growth, the latter two DPRK tests were big enough to induce a few aftershocks, which were be detected by the closest IMS as well as non-IMS stations. Two from six aftershocks (see Table 2) were detected by automatic processing system of the International data centre (IDC) of the CTBTO. Corresponding event hypotheses were automatically created and then interactively reviewed and confirmed by experienced IDC analysts. Their body wave magnitudes are 4.1 and 3.4. The second and the biggest aftershock in table 2 occurred 8.5 minutes after the DPRK6: September 3, 2017; 08:38:31.2 UTC.

Three aftershocks were detected by the IDC prototype waveform cross correlation system, which correlates continuous waveforms at four regional and a few teleseismic primary IMS stations with signals (templates) from several previously detected DPRK events. The IDC analysts also confirmed these events but the corresponding event hypotheses did not have enough weight (reliability) to be promoted into the official IDC product – the Reviewed Event Bulletin (REB), which is also published for the convenience of the broader monitoring community by the

International seismological centre (ISC) with a two-month delay. These three aftershocks were saved in the IDC database as seismic events with only 2 (the minimum number for the REB is 3) primary IMS stations associated.

Table 2. Parameters of five aftershocks. $m_b$ – relative magnitude.

| Aftershock | SEL3($m_b$) | LEB ($m_b$) | REB ($m_b$) | Date | Time | $m_b$(IDC) |
|---|---|---|---|---|---|---|
| 1 | - | + | - | 11.09.2016 | 1:50:49.86 | |
| 2 | + | + | + | 3.09.2017 | 3:38: 31.88 | 4.1 |
| 3 | - | + | - | 23.09.2017 | 4:42: 59.95 | |
| 4 | + | + | + | 23.09.2017 | 8:29: 16.29 | 3.4 |
| 5 | - | + | - | 12.10.2017 | 16:41: 8.11 | |

One event was detected by the same cross correlation procedure, but with signals from the previously detected aftershocks used as waveform templates. The waveforms from aftershocks are more similar to themselves than to the explosion signals. The difference between aftershock and explosion waveforms is also expressed in the applicability of the P/S spectral amplitude ratio as a discrimination criterion.

Seismic magnitude of the DPRK5 aftershock four from five DPRK6 aftershocks were small: from 2.4 to 3.4. One can assume that the biggest aftershocks of the other four DPRK events were also by about 2.5 units of magnitude lower that their respective main shocks, i.e. they would have magnitudes below 2.4. Therefore, the current IMS network was not capable to detect such aftershocks of the previous smaller tests in routine IDC processing as well as in the prototype cross correlation processing with the explosions as master events. We are going to process longer periods after DPRK2 thought DPRK4 using the newly found aftershocks as master events. Table 2 lists general information of five from six aftershocks obtained by standard and cross correlation methods applied by the IDC routinely. The sixth event has to be discussed separately as detected by the aftershock waveform templates.

Figure 1 shows relative positions of four seismic stations closest to the DPRK test site with data available from the IMS and/or other open sources. We used the same set of stations for the analysis of the DPRK5 aftershock [8]. Two non-IMS stations MDJ (China) and SEHB (South Korea) are closer to the DPRK test site than two IMS array stations USRK and KSRS. Figure 2 through 5 present typical seismograms measured at each four stations. The difference between frequency dependent behavior of the seismic wave-field generated by the DPRK tests and their aftershocks is well illustrated. Table 3 lists six frequency bands (FB1 to FB6), which represent a comb of octave 3-rd order Butterworth filters. The frequency bands used for the spectral amplitude estimates are slightly different. Stations KSRS and SEHB are characterized by sampling rate of 20 Hz, with the Nyquist frequency of 10 Hz. At stations USRK and MDJ, the sampling rate is 40 Hz.

The sequence of seismic phases at four stations, which are situated at almost the same epicentral distances between 3.3° and 4° (see Figure 1), is standard - $P_n$, $P_g$, $L_g$. The $S_n$ phase is not seen on these traces. At KSRS, the $P_n$-phase generated by explosions is of the highest amplitude among all phases at higher frequencies, but the $P_g$-phase is of the same amplitude. The $L_g$-phase is better seen at lower frequencies and practically disappears at high frequencies. The wave-field generated by aftershocks is different with the $L_g$-phase dominating. The $P_n$- and $P_g$–wave arrivals are barely seen at higher frequencies. At USRK, the wave-pattern is different with the $P_g$-wave dominating in P-wave group. The $L_g$-phase is the largest among all phases generated by the aftershocks. The relative phase amplitudes and spectral content of the wave-fields observed at stations SEHB and MDJ from explosions and aftershocks are similar to those at KSRS and USRK, respectively, because of similar propagation paths.

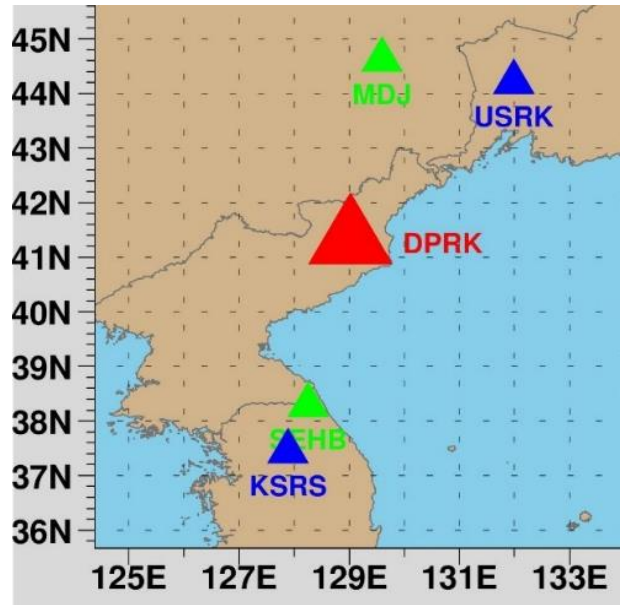

Figure 1. Regional stations used in the aftershock analysis. Blue – IMS arrays, green – non-IMS 3-C stations. DPRK test site shown by red.

Table 3. Frequency bands of the 3-rd order Butterworth filters. Notice that the high-frequency band for KSRS and SEHB is limited to 10 Hz.

| Band, Hz | KSRS | USRK | MDJ | SEHB |
|---|---|---|---|---|
| FB1 | 1-2 | 1-2 | 1-2 | 1-2 |
| FB2 | 1.5-3 | 1.5-3 | 1.5-3 | 1.5-3 |
| FB3 | 2-4 | 2-4 | 2-4 | 2-4 |
| FB4 | 3-6 | 3-6 | 3-6 | 3-6 |
| FB5 | 4-8 | 4-8 | 4-8 | 4-8 |
| FB6 | 6-10 | 6-12 | 6-12 | 6-10 |

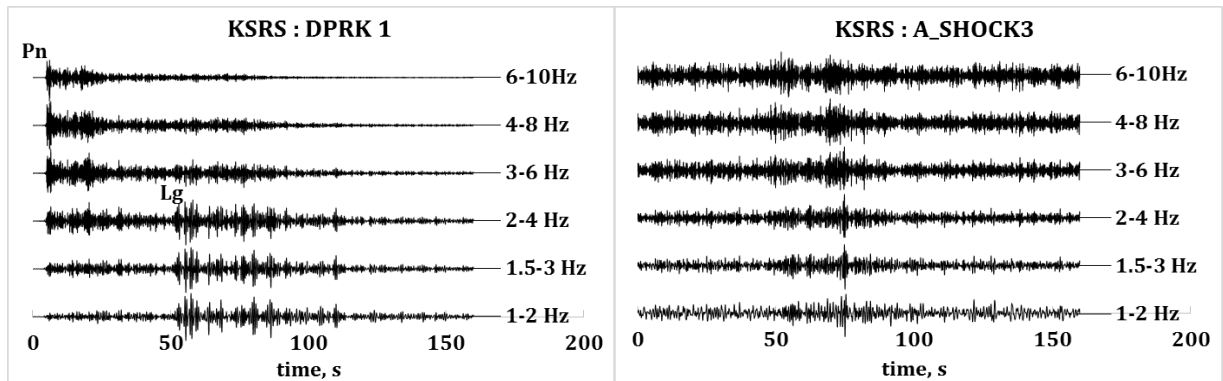

Figure 2. Slowness-adjusted beams calculated for the first and smallest explosion () and the third aftershock (23.09.2017, 4:42:59.95 UTC) at stations KSRS. All signals are filtered in 6 frequency bands (Table 3) and normalized to their respective peak values.

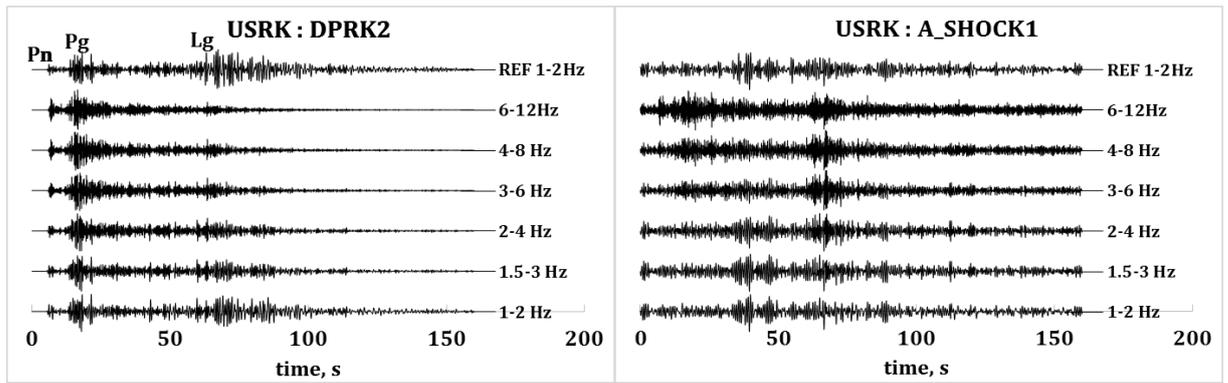

Figure 3. Same as in Figure 2 for USRK. The array reference channel filtered between 1 and 2 Hz is also shown.

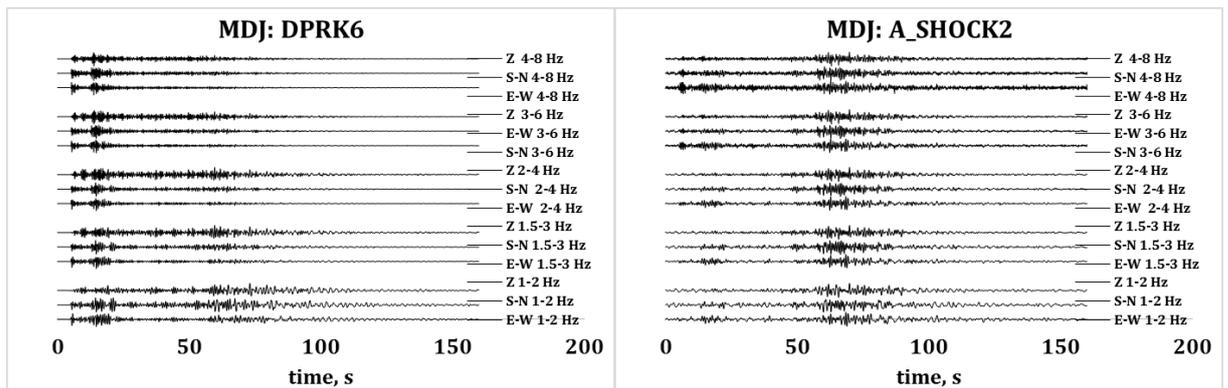

Figure 4. Comparison of 3-C filtered waveforms at station MDJ. The DPRK6 and the aftershock 2 are presented.

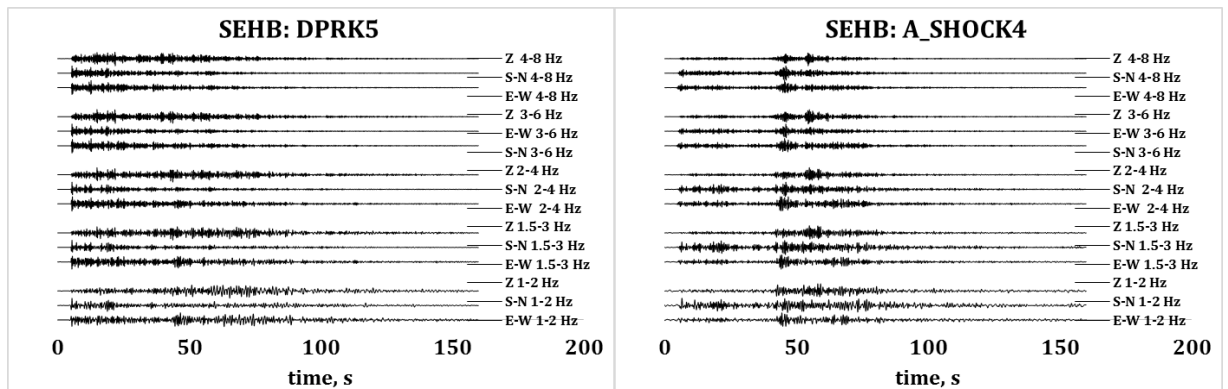

Figure 5. Same as in Figure 4 for station SEHB. The DPRK3 and the aftershock 4 are presented.

**Detection using waveform cross correlation with the aftershocks templates**

The number of events measured at 4 stations varies from 11 (6 explosions and 5 aftershocks) at KSRS to 7 at SEHB. The DPRK1 was not measured by USRK because it did not work in 2006 and no data are available for the DPRK1 at MDJ. Waveforms from only two last explosions are available at SEHB. All these signals were detected by the waveform cross correlation method with the DPRK tests used as master events, i.e. all 5 aftershocks were found by explosion templates. However, Figures 2 through 5 demonstrate that there exists a dramatic difference in the features of seismic wave-field generated by these two types of seismic events. This observation indicates that the aftershock signals could better cross correlate with each other than with the explosion signals, and thus, provide a more sensitive detector of even smaller aftershocks.

Figures 6 and 7 depict selected waveform templates at stations USRK and KSRS, respectively. These templates represent seismic signals measured at the reference array channels, which were filtered in various frequency bands since the difference in shape between the explosion and aftershock signals varies with frequency band. The total template length is 155 s with a 5 s pre-signal interval added to allow more flexibility in start time. (For arrays, the start time at different channels/sensors can be later or earlier than that on the reference channel.) The template length is enough to include all regional phases of interest. Six frequency bands in Table 3 cover the whole spectral range relevant to the studied seismic waves and allow improving signal-to-noise ratio depending on seismic phase and noise conditions at vertical and horizontal (for 3-C stations) components.

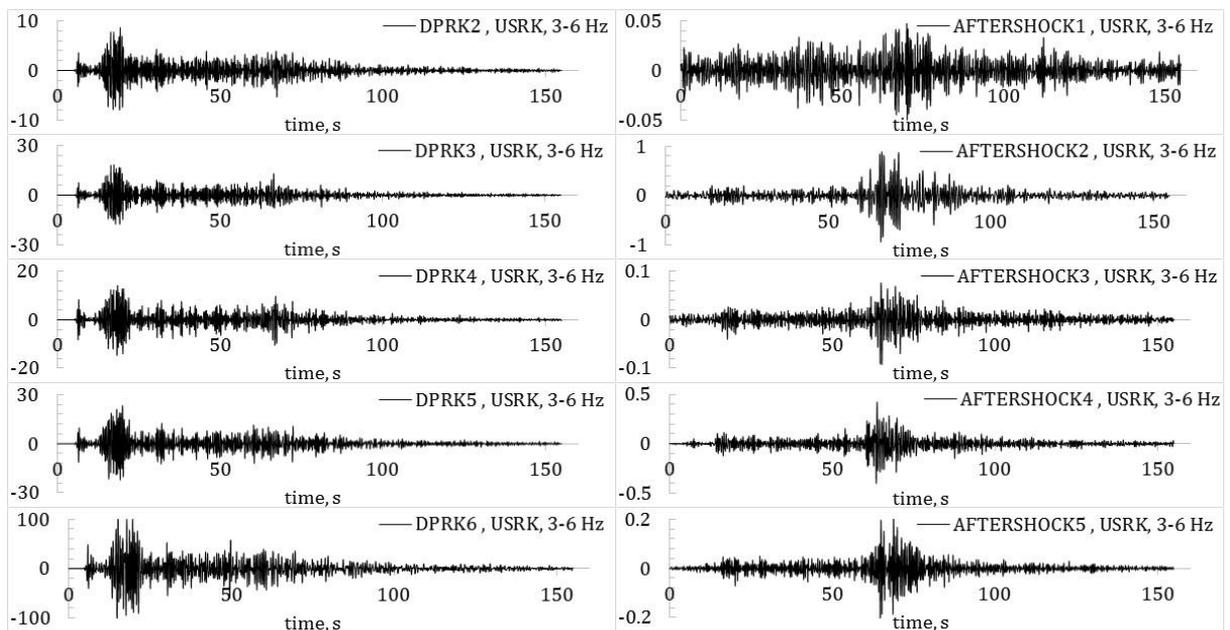

Figure 6. Ten waveform templates (left panel – explosions, right panel – aftershocks) at the reference channel in the frequency band 3 to 6 Hz for USRK.

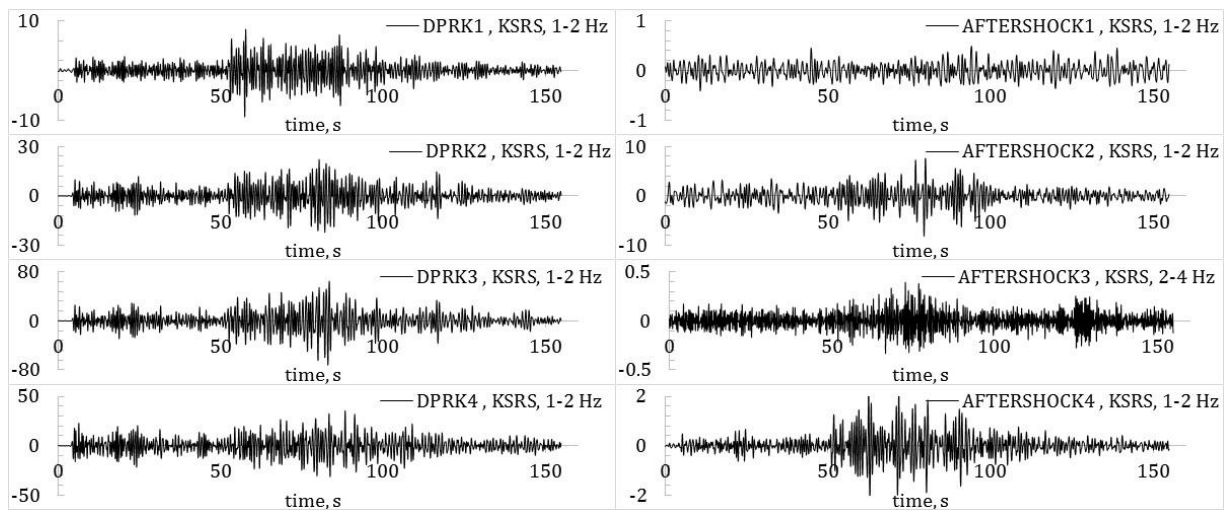

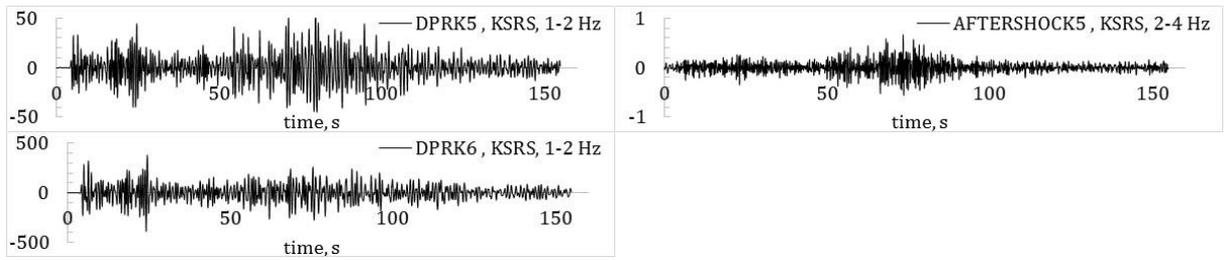

Figure 7. Waveform templates at station KSRS. Band-pass filters 1 to 2 Hz and 2 to 4 Hz.

The method of detection based on waveform cross correlation includes several steps [9-13]. At each station, all available waveform templates from the DPRK explosions and aftershocks are correlated with continuous data at corresponding individual channels separately. For arrays, we use only vertical channels, and for 3-C stations one vertical and two horizontal components are used. To obtain a single CC-trace for a station, all individual CC-traces are averaged for each discrete time point. As a result, a unique CC-trace is created with the same sample rate as the original trace. At this trace, any real signals from all seismic sources different from the master events are considered as noise and have to be suppressed by destructive interference with the waveform templates.

The effectiveness of a waveform template in detection of signals similar to those from the master events depends not only on the similarity with the sought signal from slave events, but also on the dissimilarity with the ambient noise. The latter might be of higher importance because the highest *CC* value is 1.0 for autocorrelation. At the same time, there is no low-amplitude limit to the level of noise in the CC-traces. The ratio of the signal *CC* and the average level of *CC* in the pre-signal noise defines the detector efficiency.

Figure 8 illustrates the difference between CC-traces at station KSRS obtained with the DPRK1 and the aftershock 2 (A_SHOCK2) as master events. In both cases, the template length is 140 s and the frequency band is 1.5 to 3 Hz. These are not optimal values for these templates and are selected only for illustration purposes. The length of interval is 1 hour or 72,000 discrete time points for the 20 Hz sampling rate. The signal from the aftershock 1 (A_SHOCK1) was detected at KSRS by the cross correlation method at 1:51:40 UTC. This onset time corresponds to both observed CC-peaks.

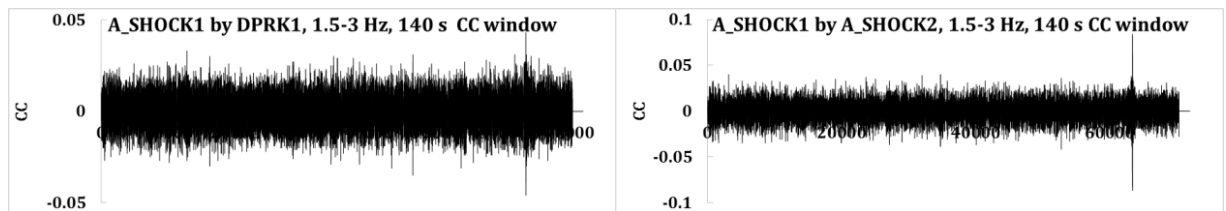

Figure 8. One-hour-long traces of the cross correlation coefficient, *CC*, at station KSRS calculated with the DPRK2 (left panel) and the second aftershock, A_SHOCK2, (right panel). Both peaks correspond to the arrival time of signals from the first aftershock (11.09.2016). The aftershock template gives a higher CC-value and a slightly higher noise level.

The CC-signals in Figure 8 are different from real seismic signals. They are short (have no coda) and sharp likely allowing accurate onset time estimates. Figure 9 shows that the highest |**CC**| from the A_SHOCK2 is approximately 2 times larger than that from the DPRK1 despite both events have the same body wave magnitude (according to the IDC estimates in Tables 1 and 2). This is likely explained by higher similarity between the template signal from the A_SHOCK2 and the sought signal from the A_SHOCK1. Figure 10 depicts corresponding $SNR_{CC}$ traces and confirms that the A_SHOCK2 template provides a more sensitive detector for the aftershock 1. There are three close peak $SNR_{CC}$-values from the A_SHOCK2. In order to determine the onset time we use the peak |**CC**| within 1 s from the highest $SNR_{CC}$ value. Considering the case in Figures 9 and 10, the third negative CC-peak (-0.087 with the highest positive value of 0.084) defines the arrival time instead of the fourth peak in the $SNR_{CC}$-trace. For the DPRK1, the arrival time is also defined by the largest in absolute amplitude negative CC-peak (-0.046) with two close peaks in the $SNR_{CC}$-trace.

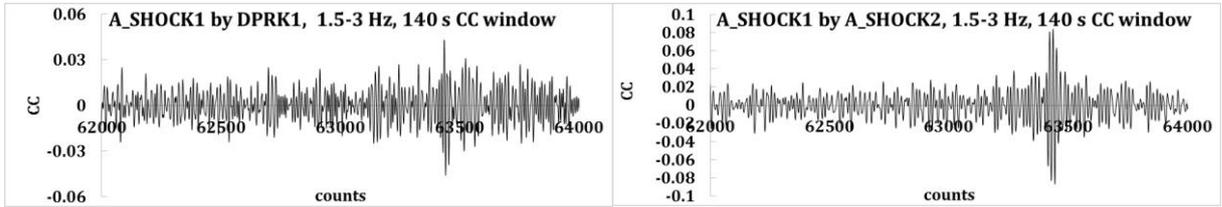

Figure 9. Two peaks shown in Figure 8. The A_SHOCK2 peak value is approximately higher by a factor of 2. The pre-signal noise for the DPRK1 is marginally lower.

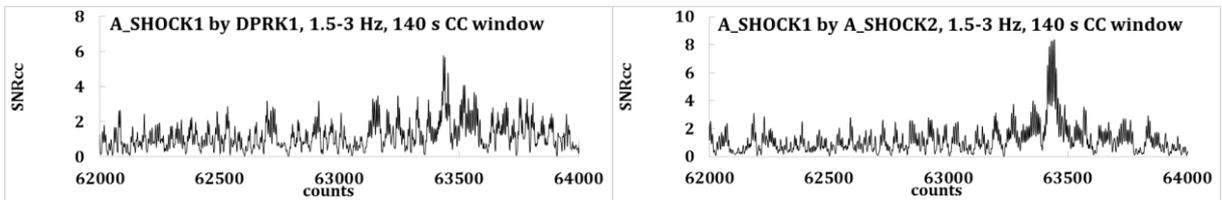

Figure 10. The $SNR_{CC}$ (STA/LTA) traces for the CC-traces in Figure 9 with STA=0.1 s and LTA=20 s. The peak $SNR_{CC}$ is larger for the A_SHOCK2 as a master event.

Having the detection pipeline based on waveform cross correlation we correlate all available signals to each other varying template length, frequency band, and defining parameters of the STA/LTA detector in order to find an optimal set of parameters to build the most effective detector of smaller aftershocks, which does not produce too many false alarms. The length of cross correlation window varied between 10 s and 150 s with a 10-second step. This allows sequential inclusion of all regional phases important for further investigation of the P/S spectral amplitude ratio. The length of the long-term-average (LTA) was always 20 s, which is a relatively short interval for real seismic signals varying in a broader range of amplitudes. However, it expresses the fact that the estimated *CC* values are always limited to the range between -1 and 1. Therefore, cross correlation of any transient signal with the ambient noise creates a rather stationary time series with the exclusion of correlation with similar signals.

At the initial stage, the length of the short-term-average (STA) window was varied between 0.1 s and 1.0 s. (We average the absolute *CC* values.) After a few runs, it was found that the 0.1 s STA provides the highest sensitivity without a large number of false alarms at the threshold level of STA/LTA=5. Therefore, STA=0.1 s does not affect the reliability of event hypotheses built using the cross correlation detection method because the consecutive arrivals obtained with a given template are separated by minutes and tens of minutes. The probability of wrong association of noise detections at 4 stations with 1 s arrival time (or origin times with known travel times for a given master event) residuals can be estimated as the product of probabilities of a random detection to be with 1 s of the associated detection at a given station. With constant probability for a noise detection (false alarm) to occur at any time and the rate of noise detections of 1 per 600 s one has $2/600 = 3.2 \times 10^{-3}$ probability of wrong association. For a 4-station event, the joint probability is $10^{-10}$. We could further reduce the detection threshold below STA/LTA=5, but at this stage we are looking for very reliable aftershocks, which could be used in the discrimination study.

All six frequency-bands occurred to be important for the CC-detector optimization because of changing ambient noise conditions and spectral content of regional phases. For example, the $L_g$-waves are characterized by relatively intensive low-frequency content and the ambient noise includes low-frequency surface waves propagating in the uppermost part of the crust, with amplitudes depending on time and weather conditions. The $P_n$ and $P_g$-waves are of much higher frequencies, where the input of industrial sources might be large depending on station position. Since we are searching for the global $SNR_{CC}$ maximum over all defining parameters, in some cases the best and second best settings may include different frequency bands. For the small aftershocks detection we use the best setting and ignore other settings even if they are very close to the best one.

Some principal results of the optimization procedure are shown in Figures 11 and 12. (Notice that for the sake of completeness these tables also include an additional aftershock (A_SHOCK6), which was found in the extended detection procedure with the aftershocks templates.) The highest $SNR_{CC}$ (=STA/LTA) obtained at KSRS vary from 5 (i.e. the threshold value) to 657, and USRK is characterized by the range between 5.3 and 539. It is instructive that in some cases the peak $SNR_{CC}$ is observed not at the matrix diagonal i.e. does not belong to auto-correlation. We have already discussed this effect and explained the importance of noise suppression by a given template.

Even a superficial visual inspection of the $SNR_{CC}$ matrices reveals the presence of a few clusters. All explosions correlate better between themselves than with the aftershocks. The DPRK1 slightly differs from other five events with much higher auto-correlation. This effect is likely related to its location at distances >2.6 km from the other DPRK explosions, which were conducted at distances less than 1.5 km from each other. Three aftershocks (A_SHOCK3 through A_SHOCK5) reveal a high level of cross correlation with each other. The auto-correlation gives the highest $SNR_{CC}$, however. This is definitely a cluster, which should express their close locations and overall similarity in seismic wave generation. The aftershock 1 is best correlating with the aftershock 6 and both (being the smallest aftershocks events) do not correlate well with other aftershocks. The A_SHOCK2 is best correlating with the A_SHOCK4, but this can be the size effect since these are two biggest aftershocks (see Table 2).

|          | D1  | D2  | D3 | D4  | D5  | D6  | A1   | A2  | A3  | A4  | A5   | A6  |
|----------|-----|-----|----|-----|-----|-----|------|-----|-----|-----|------|-----|
| DPRK1    | 221 | 73  | 45 | 34  | 50  | 66  | 8.3  | 37  | 15  | 31  | 23.1 | 5.7 |
| DPRK2    | 88  | 373 | 76 | 246 | 344 | 293 | 8.6  | 34  | 18  | 40  | 16.6 | 7.2 |
| DPRK3    | 80  | 534 | 92 | 201 | 468 | 381 | 11   | 47  | 18  | 33  | 14.4 | 5.8 |
| DPRK4    | 39  | 173 | 84 | 303 | 193 | 282 | 9.4  | 54  | 12  | 37  | 15.2 | 6.4 |
| DPRK5    | 34  | 135 | 56 | 537 | 120 | 657 | 8.7  | 42  | 14  | 35  | 17.7 | 6.5 |
| DPRK6    | 48  | 141 | 70 | 196 | 199 | 597 | 9.9  | 51  | 15  | 36  | 14.5 | 5.6 |
| A_SHOCK1 | 5.7 | 9.1 | 7.6| 7.8 | 9.4 | 6.7 | 107  | 7.3 | 8.8 | 19  | 12.4 | 23  |
| A_SHOCK2 | 30  | 53  | 29 | 59  | 72  | 72  | 11   | 112 | 12  | 27  | 14   | 9.2 |
| A_SHOCK3 | 19  | 13  | 13 | 15  | 13  | 18  | 9.2  | 10  | 87  | 144 | 97.3 | 8.5 |
| A_SHOCK4 | 21  | 65  | 30 | 24  | 81  | 30  | 19   | 21  | 59  | 209 | 79.3 | 16  |
| A_SHOCK5 | 32  | 20  | 11 | 14  | 23  | 20  | 14   | 12  | 63  | 85  | 151  | 15  |
| A_SHOCK6 | 5.2 | 6.3 | 5.7| 5.5 | 6.2 | 5.9 | 28   | 5   | 8.5 | 13  | 10.7 | 98  |

Figure 11. Maximum $SNR_{CC}$ values obtained at station KSRS in the detector optimization procedure with varying defining parameters. Colors highlight clusters. (D1=DPRK1, etc.)

The length of optimal cross correlation windows covers the studied range from 10 s to 150 s. For the DPRK explosions, the shortest template sometimes gives the highest $SNR_{CC}$ because the first 10 s are of the highest amplitude in all frequency bands. For the aftershocks, the CC-window length is much longer and mainly varies between 120 s and 150 s because the $L_g$-wave is the most intensive. Hence, we observed a significant difference in the optimal detector parameters between the explosion and aftershock groups, but there are also some differences within these groups. Since the highest $SNR_{CC}$ can be reached for any template length and any frequency band, we use the whole range of these two parameters in the routine detection procedure.

|          | D2   | D3  | D4   | D5  | D6   | A1  | A2  | A3   | A4  | A5  | A6   |
|----------|------|-----|------|-----|------|-----|-----|------|-----|-----|------|
| DPRK2    | 204  | 50  | 273  | 248 | 273  | 14  | 13  | 14   | 42  | 40  | 8.21 |
| DPRK3    | 228  | 62  | 182  | 166 | 182  | 16  | 18  | 12   | 34  | 19  | 8.58 |
| DPRK4    | 226  | 34  | 539  | 301 | 539  | 23  | 29  | 9.7  | 23  | 10  | 11.2 |
| DPRK5    | 274  | 38  | 291  | 400 | 291  | 26  | 25  | 14   | 16  | 14  | 13.3 |
| DPRK6    | 252  | 68  | 212  | 356 | 212  | 25  | 27  | 12   | 21  | 14  | 14.7 |
| A_SHOCK1 | 22   | 13  | 24   | 27  | 24   | 85  | 13  | 6.1  | 9.5 | 7.3 | 17   |
| A_SHOCK2 | 47   | 11  | 71   | 43  | 71   | 15  | 160 | 15   | 51  | 13  | 8.89 |
| A_SHOCK3 | 9.8  | 11  | 9.1  | 12  | 9.1  | 7.2 | 24  | 67   | 75  | 119 | 6.78 |
| A_SHOCK4 | 12   | 11  | 14   | 22  | 14   | 12  | 9.2 | 132  | 148 | 83  | 10.2 |
| A_SHOCK5 | 17   | 8.1 | 10   | 15  | 10   | 10  | 22  | 116  | 123 | 100 | 5    |
| A_SHOCK6 | 9.3  | 6.8 | 11   | 10  | 11   | 17  | 9.3 | 7.6  | 16  | 5.3 | 66.4 |

Figure 12. Maximum $SNR_{CC}$ values as obtained at station USRK in the detector optimization procedure with varying defining parameters.

**Association of cross correlation detection obtained by standard and combined master events**

We have developed a special procedure of association with a unique event for detections obtained by the waveform cross correlation method at several stations [11-13]. This procedure is called the *local association*, *LA*, in line with the name of global association, *GA*, used by the IDC [14]. In this procedure, only the phases from events close (local) to the master ones are associated. The *LA* should not sensitive to any events beyond the radius of correlation, which depends on the source-station distance as well as on the slowness and duration of detected seismic signals. Within the predefined correlation radius, where we are looking for a solution matching arrival times at as many stations as possible, a fine grid of potential slave locations is introduced. For its nodes, we estimated the master-station travel time corrections using standard travel time curves. In a few unusual cases, when a distant event (i.e. out of the correlation radius) demonstrates at some stations relatively high cross correlation coefficients with a given master event, the origin times, which can be obtained from the arrival times by subtraction of the master-station travel times, must scatter beyond the predefined limits. In this case, no event hypothesis can be built.

For two events with close locations, the level of cross correlation coefficient at a given seismic station (array of 3-C) depends on the distance between events, the similarity of source functions, the difference in velocity/attenuation structure along propagation paths, and the change in spectral characteristics of the ambient microseismic noise. Apparently, for larger master/slave distances *CC* should be lower, *ceteris paribus*. Together with decreasing *CC*, the slave event offset results in the travel time change. For an array station, this offset can also change the travel time delays between sensors of the array relative to the reference channel.

The work of the cross correlation detector results in a set of detections characterized by their arrival times, $at_{ij}$, where $i$ is the index of the $i$-th arrival at station $j$. For a slave close but not coinciding with the master event, the travel times to corresponding stations can be accurately approximated as the master-station travel times corrected for the master-slave coordinate difference, **d**, as estimated using, for example, the IASPEI91 global velocity model:

$$dt_{ij} = \mathbf{s}\cdot\mathbf{d} \qquad (1)$$

where $dt_{ij}$ is the travel time correction for the relative coordinates **d**, and **s** is the vector slowness estimated by the IASPEI91 model for the master-station pair. Therefore, when we create a slave event hypothesis with fixed coordinates relative to the master event, one can calculate the origin times, $ot_{ji}$, for all detections using the corrected travel times:

$$ot_{ji} = at_{ij} - tt_j + dt_{ij} \qquad (2)$$

The set of arrival times is converted into a set of origin times for an unknown number of hypothetical slave events. By definition, an origin time is the most important characteristic of any seismic source. It allows time association of several arrivals with a unique event. So far, we do not know the best slave location. Moving the slave position over the fine grid, we first estimate the number of stations with the origin times within the predefined tolerance, which is 2 s for the current study. Since we assume that all slave events have to be within a few kilometres, the grid size is 10×10 km with the master in the centre and the largest distance from it of about 7 km. The grid step is 100 m. For an event hypothesis, we average all associated origin times and assign the estimated value to the event origin time. This gives us corresponding origin time residuals, and we are looking for a location with the highest number of stations and the lowermost RMS origin time residual, the former being the first priority This is the slave location, which corresponds to the best event hypothesis.

The relative location technique works best when the origin time residuals are smaller than the time corrections in (2). Then one can obtained the relative location accuracy of a few hundred of event tens of meters, as obtained for the DPRK explosions except the DPRK6 [paper in preparation]. The RMS origin time residuals obtained for 4 regional IMS stations from the DPRK explosions are of the order of 0.01 s with the travel time corrections of the order of 0.1 s for the distance between events of 1 km and the $P_n$-wave velocity of 8 km/s. When the travel time residuals exceed 0.1 s, the relative location method fails because the causes of such residuals are not related any more to the distance between events. The absence of extremely low RMS origin time residual does not mean, however, that the slave event hypothesis loses its reliability. The latter is defined by the probability of a false event creation by random detections, as was discussed before.

We use all 11 master events separately and have to choose which slave hypothesis is the best when more than one master detects the same slave event. The conflict between competing hypotheses is resolved by the choice of the event with the smallest RMS origin time residual. The best event may have lower **SNR**$_{CC}$ at some or all involved stations. After continuous cross correlation of 11 masters between September 3 and November 30, 2017 we found only one additional aftershock event. Table 5 lists the origin time, arrival times at 4 stations, travel time residuals, *SNR*$_{CC}$ and standard *SNR*. This event occurred on September 3, 2017 at 09:31:30 UTC. It was missed in standard detection processing. The low *SNR* values explain this result. It was also missed by the prototype cross correlation processing. This failure is partially explained by the fact that the best A_SHOCK6 hypothesis was created by the A_SHOCK4, which gave very low travel time residuals. Figure 13 depicts two CC-traces obtained with A_SHOCK1 and A_SHOCK4 as master events. Both peaks are very sharp and thus give accurate arrival time estimates. In line with Figure 11, the A_SHOCK1 gives the largest $SNR_{CC}$ value for the A_SHOCK6, and the A_SHOCK4 is the second best. The first aftershock after the DPRK5 (A_SHOCK1) was too small to create quality waveform templates at all 4 stations and it failed as the best master.

Table 5. Solution for the aftershock 6 with the best master (aftershock 4). Origin time 09:31:30 UTC.

| Station | Arrival time | Residual, s | $SNR_{CC}$ | SNR |
|---|---|---|---|---|
| **SEHB** | 09:32 :20.87 | 0.37 | **7.3** | 2.6 |
| **MDJ** | 09:32 :22.34 | 0.03 | **6.1** | 4 |
| **USRK** | 09:32 :25.98 | -0.02 | **5.1** | 4.8 |
| **KSRS** | 09:32 :32.3 | -0.37 | **7** | 5.5 |

There are several solutions for the A_SHOCK6 created by other aftershocks and even more CC-detections found by the DPRK explosions. Obviously, additional 4-station solutions for the same slave event (*e.g.*, A_SHOCK6) indicate that this event is more reliable than the event found by just one master. In general, more detections we have from all masters, which can be associated with an event with the same origin time, more reliable is the corresponding event hypothesis. We call joint usage of all master events for creation of a unique event hypothesis – the combined master event. This approach is applicable when all master events are very close to each other and their templates give the same arrival time at a given station. The DPRK explosions and aftershocks are all within 3 to 4 km from each other. The results of cross correlation detection show that the arrival times obtained by different masters (templates) may differ by a second or so for a given slave. For the DPRK explosions as masters and slaves, this difference is much smaller, however. In order to synchronize all detections we selected the DPRK5 as a reference event with fixed travel times to all stations. Then the difference between the arrival time obtained by auto-correlation (i.e., DPRK5) and other events is calculated. For the combined master, all arrival times are corrected for the obtained differences and the DPRK5 travel times are used in the *LA* process.

Table 6 lists the solution obtained by this combined master consisting of 11 individual masters. In total, there are 25 associated detections (from 38 different templets) within 2 s from the average origin time 9:31:29.86 UTC. The not-detecting templates are related to the DPRK explosions. Chiefly, the origin time residual are lower than 0.5 s, with only 3 residuals larger than 1 s. Input of stations vary from 4 (from 7) for SEHB to 9 (from 10) for USRK. Station KSRS gives 7 associated detections from 11 masters. The combined-master solution has some $SNR_{CC}$ values much higher than those in the single-master one. Overall, the larger number of associated arrivals with larger $SNR_{CC}$ makes this seismic event very reliable.

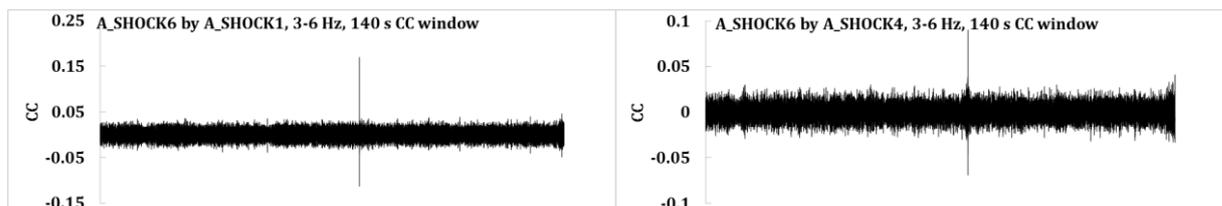

Figure 13. CC-traces at KSRS obtained with the A_SHOCK1 and A_SHOCK4 as master events. Two sharp CC-peaks correspond to the same signal generated by this aftershock.

Table 6. Joint solution for the aftershock 6. Date 3.09.2017, origin time 9:31:29.86 UTC.

| Station | Time | Residual, s | SNR$_{CC}$ | SNR |
|---|---|---|---|---|
| SEHB | 09:32 :18.64 | -1.13 | **43.6** | 3.4 |
| SEHB | 09:32 :19.99 | 0.22 | **21.9** | 2.3 |
| SEHB | 09:32 :20.69 | 0.92 | **6.2** | 3.2 |
| SEHB | 09:32 :21.44 | 1.67 | **7.3** | 2.6 |
| MDJ | 09:32 :22.59 | -0.60 | **5.8** | 4.0 |
| MDJ | 09:32 :23.19 | 0.00 | **5.5** | 3.9 |
| MDJ | 09:32 :23.41 | 0.22 | **48.8** | 3.9 |
| MDJ | 09:32 :23.53 | 0.35 | **6.1** | 4.0 |
| MDJ | 09:32 :23.61 | 0.43 | **5.9** | 4.0 |
| USRK | 09:32 :25.28 | -0.69 | **15.7** | 4.8 |
| USRK | 09:32 :25.50 | -0.47 | **8.0** | 4.8 |
| USRK | 09:32 :25.55 | -0.42 | **13.0** | 4.8 |
| USRK | 09:32 :25.68 | -0.29 | **7.7** | 4.8 |
| USRK | 09:32 :25.73 | -0.24 | **9.6** | 4.8 |
| USRK | 09:32 :25.88 | -0.41 | **57.9** | 4.8 |
| USRK | 09:32 :26.12 | 0.16 | **14.7** | 4.8 |
| USRK | 09:32 :26.58 | 0.61 | **5.1** | 4.8 |
| USRK | 09:32 :27.32 | 1.36 | **9.8** | 4.8 |
| KSRS | 09:32 :31.00 | -1.00 | **5.3** | 5.5 |
| KSRS | 09:32 :31.20 | -0.80 | **21.8** | 5.5 |
| KSRS | 09:32 :31.35 | -0.44 | **98.0** | 5.5 |
| KSRS | 09:32 :31.65 | -0.35 | **5.3** | 5.5 |
| KSRS | 09:32 :32.25 | 0.25 | **15.5** | 5.5 |
| KSRS | 09:32 :32.25 | 0.25 | **7.0** | 5.5 |
| KSRS | 09:32 :32.40 | 0.40 | **13.1** | 5.5 |

For the events within the DPRK test site, the combined-master cross correlation method demonstrates its superiority over the single-master approach. It allows further reduction in the magnitude of detectable aftershocks. There is no need to have even one 4-station single-master solution in the combined-master solution. The latter can gather detections from various masters. However, one has to set statistically justified thresholds to the number of different stations in the final event hypothesis and the total number of associated phases. We use preliminary values of 4 (from 4) different stations and 10 associated detections. Because of varying noise conditions, we do not demand all stations to give equal input to the final event hypothesis. In addition, we introduce station-specific weights in the combined solution, which are related to the probability of detection by a given station. Array stations KSRS and USRK have slightly larger weights of 1.5 because the false alarm rate at these stations is approximately 2 lower than that at 3-C stations SEHB and MDJ (weight 1.0). The total weight of a valid event hypothesis has to be 13, i.e. a solution with 10 associated phases has to include 6 detections from KSRS and USRK and 4 from SEHB and MDJ. In some detection runs, we have to lower these thresholds because of data availability. We also re-process the whole studied period (1 month from September 9, 2016 and 3 months since September 3, 2017) with KSRS and USRK only in order to tune the IDC prototype cross correlation pipeline. It was found, that the results of the full and the IDC-only CC-detection and Local Association are not different in terms of the number of event hypotheses. In that sense, non-IMS stations SEHB and MDJ make solutions more reliable but do no reduce the detection magnitude threshold. For further analysis, the newly found aftershock has extended the population of aftershock by one event.

There is another small event reported in [15,16], which occurred on 12 May 2012 and was detected by a temporary seismic profile in China at near-regional distances. This event was detected and located using the cross correlation methods with waveform templates from the DPRK2. When open GSN stations at regional stations (*e.g.* MDJ) were used, no detection from this event was found [17]. We have also investigated the whole April/May 2010 period with the cross correlation method using all available data from 5 DPRK events, excluding the DPRK6. No evidence of any seismic event near the DPRK test site with waveforms similar to those from the DPRK tests was found at IMS array stations KSRS and USRK.

The combined-master method allows additional reduction in detection threshold and we repeated our cross correlation processing with 6 aftershocks as master events. Application of the aftershocks should be also more efficient since Kim *et al*. [16] showed that this event has lower probability to belong to the population of explosions, i.e. it is rather a natural event, which can be similar to the DPRK aftershocks. Extensive processing with lowered detection thresholds at KSRS and USRK demonstrated no signs of the 2010 event. The aftershock 6 has relative magnitude of 2.4 and, judging by $SNR_{CC}$ values in Table 6, one can expect the detection threshold reduction by one unit of magnitude for an event like the A_SHOCK. Therefore, the detection threshold for the 2010 event has to be around **$m_b$**=1.5.

**Measuring the aftershock relative size**

Before we start the P/S spectral ratio analysis, it is important to estimate relative sizes of all events. For explosions, the estimates of body wave magnitude are available. However, only two aftershocks have **$m_b$** estimates based on teleseismic P-wave arrivals. The relative size of all six aftershocks can be better estimated using the method of $L_g$ amplitude scaling since the $L_g$-wave has the largest amplitude among all other regional phases generated by the aftershocks. For all 4 stations, we have calculated the RMS amplitudes in six frequency bands. Two curves for KSRS and USRK are shown in Figure 14. Overall, the RMS amplitude curves demonstrate synchronous fall with frequency, with the only exception of the A_SHOCK2 curve at KSRS, which falls much faster than five other curves. The cause of such behavior is not clear, but we retain in mind that the $L_g$-wave generated by this aftershock is merged with the coda waves of the DPRK6.

The relative magnitude of an aftershock can be defined as the logarithm of RMS amplitudes in the same time window and frequency band. We nominate the A_SHOCK2 with the largest **$m_b$** magnitude among all aftershocks as a reference events. The best frequency band is 1 to 2 Hz as having the largest RMS amplitudes. Table 7 lists the relative magnitudes estimated for all 4 stations. For **$m_b$**(A_SHOCK2)=4.11, the A_SHOCK6 network average magnitude is **$m_b$**=2.38±0.16. The A_SHOCK4 has **$m_b$**=3.61±0.11 and the IDC **$m_b$** estimate in Table 2 is 3.4. This difference is just marginal. The A_SHOCK1 and A_SHOCK4 have close relative **$m_b$** estimates: 2.82 and 2.95, respectively.

Therefore, we can find as small events as **$m_b$**=2.4 and even lower judging by the number of detecting templates and their $SNR_{cc}$. We have carried out an extensive search for smaller events using the combined-master approach and found no more so far.

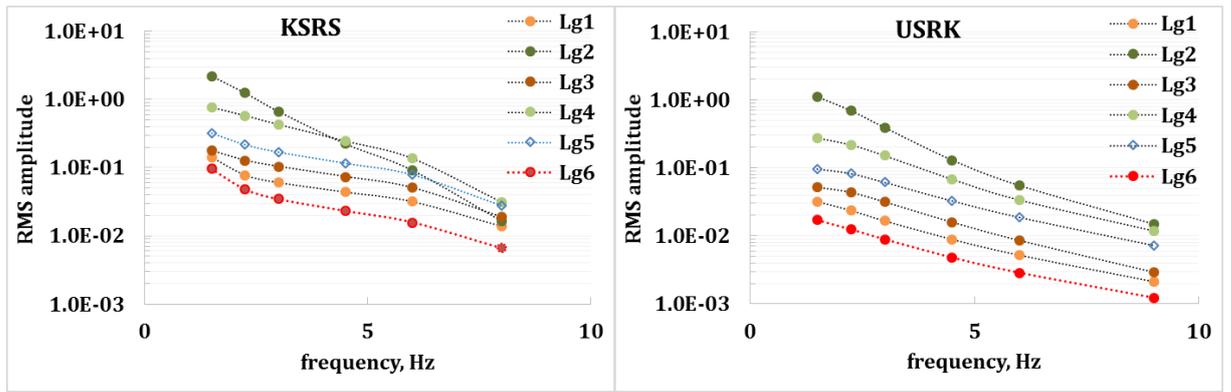

Figure 14. Frequency dependent $L_g$-wave RMS amplitude measured at KSRS and USRK.

**P/S spectral amplitude ratio as a discrimination method**

To estimate the frequency dependent P/S spectral amplitude ratios we have first to select time windows best representing the groups of P and S waves. We divide 150-s records in 10-second-long segments and calculate the RMS amplitudes in each segment in all frequency bands. For arrays, we first estimate the RMS amplitude using all individual traces. Then we steer individual traces to the reference channel with the slowness corresponding to the DPRK test site creating a single beam trace to calculate the RMS amplitude. One problem with such an approach is that group and phase velocity in the secondary phases are prone to changes relative to those of the primary $P_n$-wave. To match the changing slowness, we calculate beams with slownesses adjusted to corresponding seismic phase along the trace, i.e. the $P_n$-wave slowness for $P_n$, the $S_n$ slowness for $S_n$, and the $L_g$ slowness for $L_g$. Both approaches give approximately the same result, however, and here we use the averaging over all individual traces. For 3-C stations, we calculate two versions: only Z-component and three. No large difference was observed as well and the 3-C version is used.

Table 7. Logarithms of the RMS amplitudes divided by the RMS amplitude of the A_SHOCK2 and the network average relative body wave magnitudes of six aftershocks.

| EVENT | USRK | KSRS | MDJ | SEHB | Mean | Stdev | Relative $m_b$ |
|---|---|---|---|---|---|---|---|
| A_SHOCK1 | -1.45 | -1.14 | -1.37 | -1.19 | -1.29 | 0.13 | 2.82 |
| A_SHOCK2 | 0.00 | 0.00 | 0.00 | 0.00 | 0.00 | 0.00 | 4.11 |
| A_SHOCK3 | -1.20 | -0.96 | -1.31 | -1.18 | -1.16 | 0.13 | 2.95 |
| A_SHOCK4 | -0.50 | -0.33 | -0.64 | -0.54 | -0.50 | 0.11 | 3.61 |
| A_SHOCK5 | -0.93 | -0.72 | -0.98 | -0.91 | -0.89 | 0.10 | 3.23 |
| A_SHOCK6 | -1.73 | -1.35 | -1.64 | -1.73 | -1.61 | 0.16 | 2.38 |

Having the RMS amplitude estimates, which we use as the estimates of spectral amplitude in a given frequency band, in 15 time windows we have to select two windows with the highest RMS amplitudes for P- and S-wave. The largest difference between the explosion and aftershock is observed when the second (10-20 s) and the sixth (60-70 s) segments are used. The regional phases representing P- and S-wave groups are the $P_g$- and $L_g$-wave, respectively. For the selected P and S time windows, we have estimated the spectral amplitude ratios at four stations and Figure 15 depicts the obtained results. The number of explosions recorded by stations are 6 (KSRS), 5 (USRK and MDJ), and 2 (SEHB). The number of measured aftershocks is 6 at all stations. For the explosions, the overall behavior of the frequency dependent ratios is characterized by gradual growth. The aftershock curves usually retain their level with increasing frequency. As a result, the gap between the explosion and aftershock curves has a general tendency to increase with frequency.

Notice the *lin-log* scale in Figure 15, which results in a higher visual scattering in the aftershock P/S ratio population. At the same time, the scattering in the explosion population is larger in absolute units. Figure 16 displays two curves of mean values as obtained by averaging of all individual P/S ratios for a given source type at a given frequency for KSRS and USRK. The error bars for each mean value represent standard deviations for a given station/source-type/frequency-band combination. It is clear that the mean value and standard deviation obtained for the explosions prohibits their wrong interpretation as aftershocks. At the same time, the explosion P/S ratios at a given frequency have higher scattering, and thus, the probability to misinterpret aftershock as an explosion is higher.

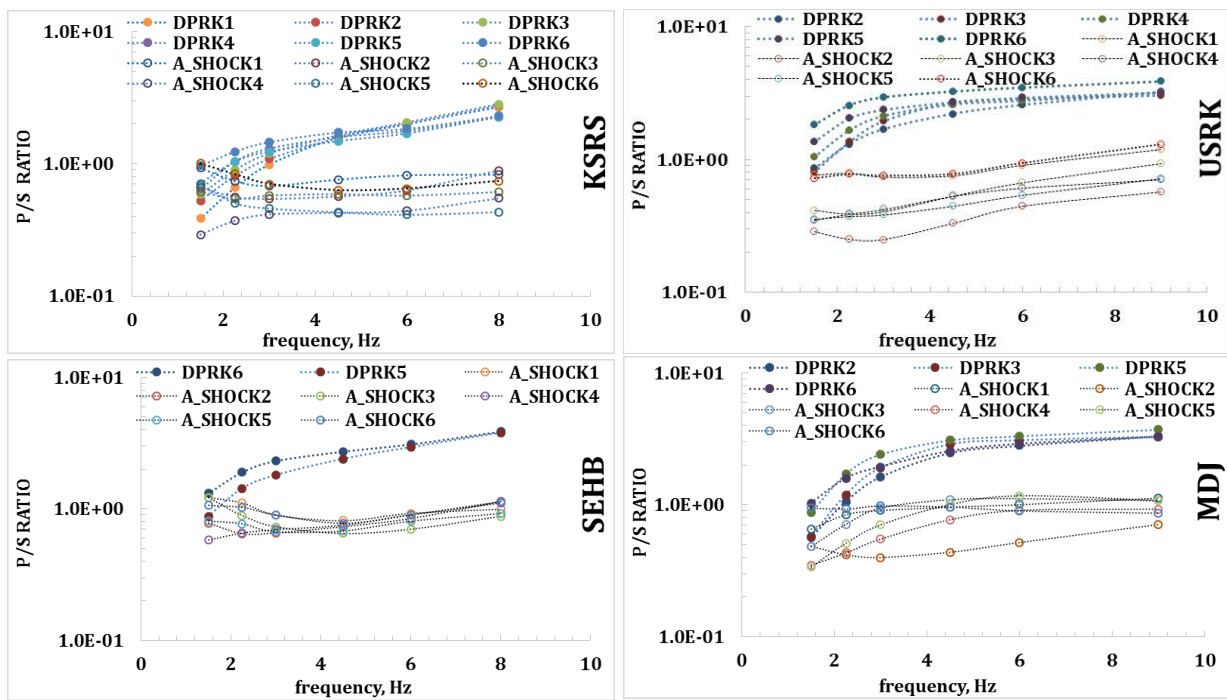

Figure 15. Dependence of the P/S ratio on frequency for stations KSRS, USRK, SEHB, and MDJ. Notice the *lin-log* scale.

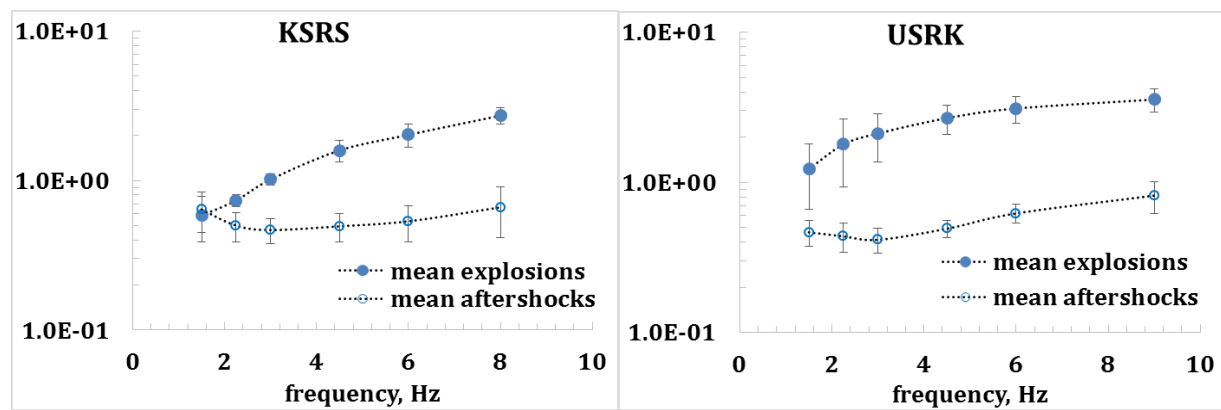

Figure 16. The mean values of the P/S ratio for the aftershocks (open circles) and explosions (solid circles) together with the standard deviations calculated for each frequency band. Left panel: station KSRS, where 6 aftershocks and 6 explosions were measured. Right panel: station USRK with 5 DPRK explosions and 6 aftershocks.

Statistically, this statement based on a small number of events, however. However, it is possible to estimate the probability of misinterpretation (discrimination) quantitatively. Figure 17 depicts the absolute difference between the mean values for explosions and aftershocks and the standard deviations from Figure 16. The difference of mean values increases with frequency at both stations with the peak value of 2.1 at KSRS and 2.7 at USRK. The standard deviation curves are close at KSRS, but the aftershock one is 3 times lower above 3 Hz. At USRK, the stdev curve for the explosions has a shelf of 0.6 above 3 Hz with *stdev* <0.1 for the aftershocks.

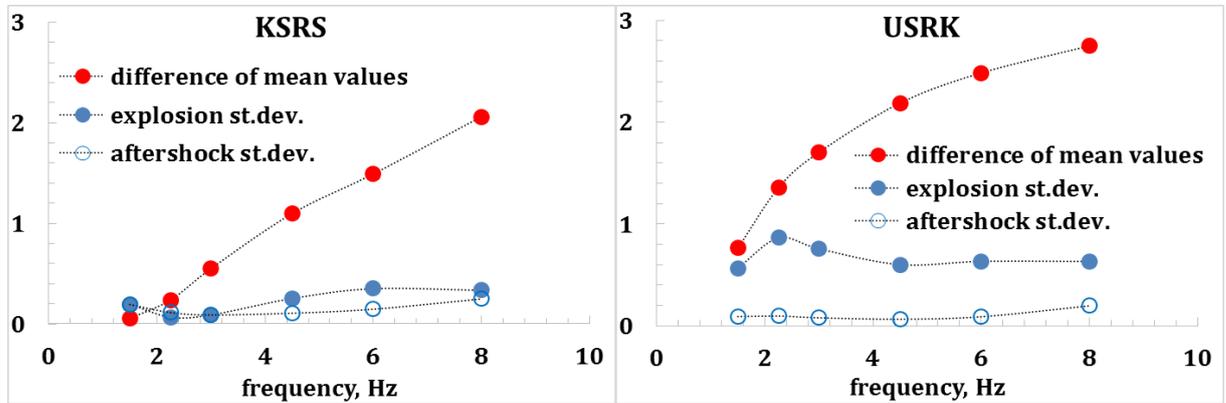

Figure 17. The absolute difference between mean values for the explosions and aftershocks and the standard deviations from Figure 16.

From Figurer 17, it is possible to estimate the distance between the mean curves belonging to two source type in corresponding standard deviations. In Figure 18, we display the ratio of the difference between the mean P/S spectral ratios for a given frequency and the standard deviation at the same frequency as obtained from the explosions and aftershocks. For USRK, the largest difference between the mean values is ~4.3 times larger than the standard deviation for aftershocks and by a factor of 33 larger than the *stdev* for the aftershocks. For the KSRS, these values are 6 and 10, respectively. The 4.3-sigma difference observed for the explosions at USRK indicates the probability of 99.996% of an average aftershock not to be an explosion. The probability of an average explosion not to be an aftershock from the observed USRK population is defined by 33-sigma, which is an infinitesimal value. For KSRS, the probability of misinterpretation defined by 6-sigma and 10-sigma. All estimated probabilities at two stations are extremely low, but we have to retain in mind that both populations are small.

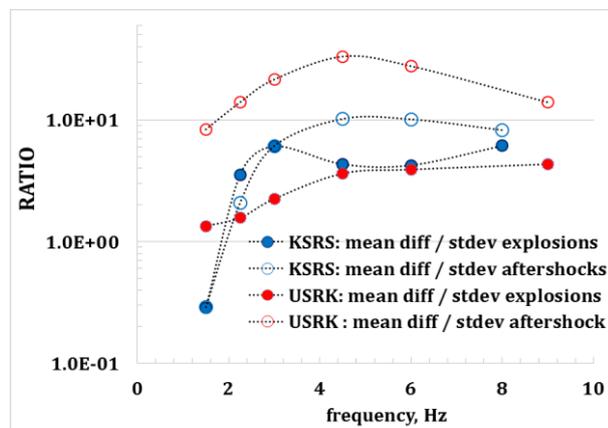

Figure 18. Ratio of the difference between the mean P/S spectral ratios for a given frequency and the standard deviation obtained from explosions and aftershocks.

**Network-wide discrimination**

In Figure 15, we presented 3 stations with the P/S estimates available from 5 DPRK tests and 6 aftershocks. At station SEHB, only two last explosions were measured. Therefore, we have independent measurements at 3 stations which can be used for network analysis of the differences between P/S ratios in the explosion and aftershock populations. The Mahalanobis distance (MD) is one of common measures in multivariate statistics to assess whether a sample is a member of a given group, an outlier or a member of a different group. This method is also effective when the distance between two populations related to different source types is the same, but they are characterized by different second statistical moments. The Mahalanobis distance depends on the internal scattering in the group and uses covariance matrix to estimate the distance between a point and a population. This is a multi-dimensional representation of the distance between a given P/S ratio and the mean P/S ratio measured in standard deviations as discussed in the previous section. The MD naturally accommodates the network-wide information.

We use the P/S spectral amplitude ratio estimates shown in Figure 15. Table 8 lists all P/S ratios used to calculate the MDs from the explosion P/S average value to six aftershocks and the MDs in opposite direction. Because of varying scattering, and thus, standard deviations within each group and between frequency bands (FBs) the MDs are quite different in two directions and change with frequency. To better characterize the reliability of discrimination between two groups, we have also to estimate the MDs for each individual member of a given group (*e.g.*, explosion) to the corresponding group average value and then compare them to the MDs to the members in another group. The difference between the internal and external MDs demonstrate the probability of a member of the alien group to be a member of the home group. We used standard Matlab application "**mahal**" to calculate Mahalanobis distance in squared units, $D^2$. For 3 degrees of freedom (3 stations) and a critical level of significance α=0.001 for the chi-square distribution, one can use the critical $D^2$=16.2. When the Mahalanobis distance measured for a given P/S spectral ratio from the group (*i.e.* explosions or aftershocks) average is larger than 16.2 one can conclude that this ratio is an outlier for this group.

Table 8. Estimates of P/S spectral amplitude ratio at 3 stations and in 6 frequency bands.

| STATION | | DPRK2 | DPRK3 | DPRK4 | DPRK5 | DPRK6 | A1 | A2 | A3 | A4 | A5 | A6 |
|---|---|---|---|---|---|---|---|---|---|---|---|---|
| KSRS | *FB1* | 0.526 | 0.587 | 0.643 | 0.713 | 0.966 | 0.928 | 0.659 | 0.615 | 0.290 | 0.702 | 1.003 |
| | *FB2* | 0.807 | 0.896 | 1.038 | 1.039 | 1.235 | 0.741 | 0.562 | 0.542 | 0.371 | 0.504 | 0.835 |
| | *FB3* | 1.089 | 1.198 | 1.302 | 1.236 | 1.459 | 0.680 | 0.543 | 0.575 | 0.414 | 0.460 | 0.703 |
| | *FB4* | 1.568 | 1.643 | 1.581 | 1.480 | 1.723 | 0.760 | 0.568 | 0.589 | 0.426 | 0.430 | 0.630 |
| | *FB5* | 1.997 | 2.056 | 1.766 | 1.694 | 1.836 | 0.819 | 0.630 | 0.575 | 0.441 | 0.412 | 0.648 |
| | *FB6* | 2.767 | 2.821 | 2.251 | 2.256 | 2.295 | 0.829 | 0.887 | 0.610 | 0.549 | 0.432 | 0.741 |
| USRK | *FB1* | 0.851 | 0.809 | 1.042 | 1.353 | 1.809 | 0.763 | 0.285 | 0.411 | 0.351 | 0.354 | 0.719 |
| | *FB2* | 1.302 | 1.355 | 1.649 | 2.033 | 2.548 | 0.786 | 0.250 | 0.385 | 0.390 | 0.372 | 0.772 |
| | *FB3* | 1.684 | 1.953 | 2.123 | 2.370 | 2.936 | 0.737 | 0.249 | 0.402 | 0.422 | 0.382 | 0.755 |
| | *FB4* | 2.196 | 2.688 | 2.589 | 2.705 | 3.241 | 0.753 | 0.329 | 0.527 | 0.530 | 0.445 | 0.777 |
| | *FB5* | 2.559 | 2.861 | 2.768 | 2.925 | 3.475 | 0.904 | 0.443 | 0.667 | 0.612 | 0.538 | 0.938 |
| | *FB6* | 3.211 | 3.036 | 3.171 | 3.205 | 3.870 | 1.190 | 0.568 | 0.933 | 0.710 | 0.718 | 1.298 |
| MDJ | *FB1* | 0.564 | 0.574 | 0.808 | 0.873 | 1.029 | 0.654 | 0.485 | 0.480 | 0.347 | 0.336 | 0.940 |
| | *FB2* | 1.059 | 1.185 | 0.931 | 1.713 | 1.589 | 0.842 | 0.417 | 0.711 | 0.432 | 0.511 | 0.884 |
| | *FB3* | 1.630 | 1.901 | 0.997 | 2.410 | 1.940 | 0.906 | 0.399 | 0.946 | 0.549 | 0.705 | 0.795 |
| | *FB4* | 2.477 | 2.870 | 1.171 | 3.092 | 2.549 | 0.962 | 0.436 | 1.096 | 0.771 | 1.007 | 0.734 |
| | *FB5* | 2.814 | 3.081 | 1.234 | 3.318 | 2.891 | 1.005 | 0.517 | 1.118 | 0.907 | 1.175 | 0.811 |
| | *FB6* | 3.312 | 3.284 | 1.177 | 3.723 | 3.279 | 1.127 | 0.709 | 1.075 | 0.928 | 1.087 | 0.952 |

In Figure 19, we present the estimates of Mahalanobis distances calculated using P/S ratios at stations KSRS, USRK, and MDJ from 5 DPRK test (2 to 6) and 6 aftershocks. All 6 frequency bands are presented in order to illustrate the performance of the Mahalanobis distance discriminator in various frequency bands. In the left panel, the MDs to the average explosion are presented. Notice the logarithm MD scale. The MDs the aftershocks to the mean explosion vary between frequency bands and between aftershocks. The MDs from 5 explosions to their average value are all between 0.46 and 3.2 independent on frequency band. The DPRK6 is characterized by the largest MDs. There are no outliers in the explosion group. The MDs from the aftershocks to the average explosion critically depend on frequency band. All 6 aftershocks have MD lower than 16.2 in one of two low frequency bands (1-2Hz and 1.5-3 Hz). For higher frequencies, there is no case with MD<23 (A_SHOCK6 in FB3). In the FB4 the smallest MD=423.6, and in the FB5 $D^2$>620. Finally, in the FB6, all MDs are between 21,000 and 27,000, i.e. by a factor of 7000 larger than the MDs from any of 5 explosions to their average value. Hence, using the high-frequency P/S spectral amplitude ratios one can prove in statistical terms that the aftershocks do not belong to the population of explosions.

In the right panel, we depict the MDs to the average aftershock. The MDs from 6 aftershocks to their average aftershock are between 0.1 and 4.2. In the FB1, the MDs from 5 explosions to the mean aftershock vary between 10 (DPRK3) and 90 (DPRK6). In other FBs, there is no $D^2$<20. The most efficient discrimination between explosions and aftershocks is achieved in the FB6 where the smallest MD=547. The estimated MD values enforce the conclusion that it is rather impossible to confuse any of the DPRK explosions with their aftershocks.

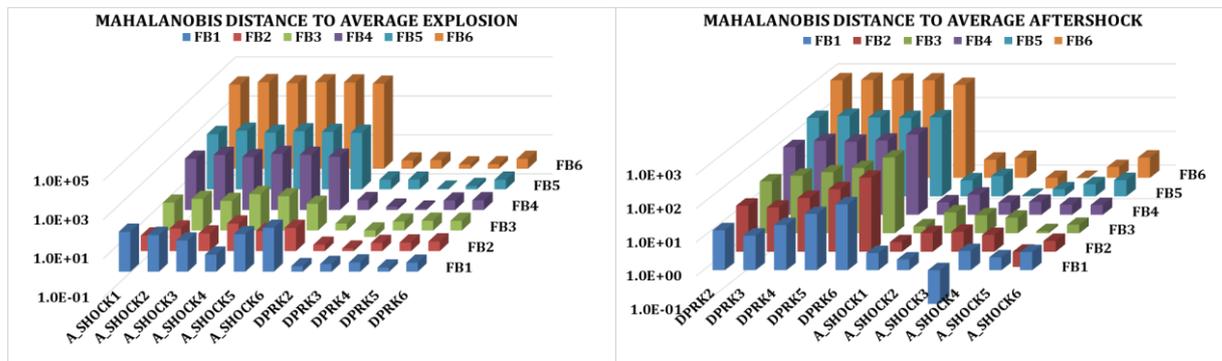

Figure 19. Squared Mahalanobis distances (MDs) calculated using the P/S ratios at stations KSRS, USRK, and MDJ from 5 DPRK test (2 to 6) and 6 aftershocks. Six frequency bands are presented. Left panel: MDs to the average explosion. Notice the logarithm scale. Right panel: MDs to the average aftershock.

**Conclusion**

We have estimated the performance of discrimination criterion based on the P/S spectral amplitude ratios obtained from six underground tests conducted by North Korea and six aftershocks induced by the last two explosions. Two seismic arrays USRK and KSRS of IMS and two non-IMS 3-C stations SEHB (South Korea) and MDJ (China) were used. The largest P and S phases at all stations ranged between 3.3 and 4 degrees were $P_g$ and $L_g$, respectively. The DPRK tests generated large-amplitude seismic waves well detected at regional and teleseismic stations. The aftershocks were so weak that advanced detection procedures were developed to detect and analyze their seismic signals.

Two from six aftershocks were detected in routine processing at the IDC. They had body wave magnitudes of 4.1 and 3.4. One can consider the lower magnitude of 3.4 as a detection threshold for standard IDC processing of the DPRK related events. Three aftershocks were found by a prototype waveform cross correlation procedure with the DPRK explosions as master events. The relative magnitudes estimated by the $L_g$ RMS amplitude scaling were between 2.85 and 3.23. Thus, the cross correlation method with the DPRK explosion templates allows reduction in the detection threshold to 2.8, i.e. 0.8 units of magnitude lower than in standard IDC processing. These five aftershocks were reviewed by IDC analysts, who confirmed them as valid events. Finally, one aftershock was found with the other five aftershocks as master event and had magnitude 2.4.

The $SNR_{CC}$ values estimated for the smallest aftershock were high enough for further reduction in the detection threshold to magnitude 1.5. We have attempted to find more aftershocks in one month period after the DPRK5 and three month period after DPRK6 with the aftershock and explosion templates combined in one master event. There was no other aftershock detected at the magnitude level of 1.5. We have also processed with the combined master event the period April/May 2010 where a very small event near the DPRK test site was reported by other researches. No event was detected in April/May 2010.

We characterize similarity of two signals in terms of cross correlation by $SNR_{CC}$, which is standard $SNR$, but measured at CC-traces. This value varies in a very broad range from 5 (threshold) to 657 for various pairs of DPRK explosions and aftershocks. By compiling a matrix of $SNR_{CC}$ values, we revealed several clusters of higher correlation, which is interpreted as source similarity or/and very short distance between events. Six explosions are characterized by extremely high correlation with each other. However, the DPRK1 is slightly different with much higher auto-correlation than correlation with the other event. We explain this effect as associated with its location at distances >2.6 km from the other DPRK explosions separated by not more than 1.5 km.

Clustering of explosion is not surprising. We have also revealed that three aftershocks (A_SHOCK3 through A_SHOCK5) are characterized by higher cross correlation with each other. This is definitely a cluster, which should express their close locations and overall similarity in seismic wave generation. The aftershock 1 is best correlating with the aftershock 6 and both (being the smallest in magnitude) do not correlate well with other aftershocks. The biggest A_SHOCK2 is best correlating with the A_SHOCK4, but this can be the size effect. Otherwise, it seems to be a standalone event 8.5 minute after the biggest DPRK6. Overall, the presence of clusters in the aftershock population creates new opportunities for investigation of source mechanisms and evolution of post-seismic activity.

The P/S spectral amplitude ratio is a well-known discriminator of explosions and earthquakes at regional distances. We have calculated P/S spectral ratios for each of four stations in six different frequency bands. With increasing frequency, all stations demonstrate approximately the same level of deviation between the $P_g/L_g$ spectral amplitude ratios belonging to the DPRK explosions and their aftershocks. For a single station, simple statistical estimates show that the probability of any of six aftershocks not to be a sample from the explosion population is larger than 99.996% at the KSRS and even larger at USRK. The probability of any of the DPRK explosion to be a representative of the aftershock population is extremely small as defined by the distance of 20 and more standard deviations to the mean explosion $P_g/L_g$ value. Station MDJ shows similar discrimination results. Station SEHB has only two DPRK tests measured, and thus, statistical estimates would be unreliable despite the populations of explosions and aftershocks at this station are well separated.

Joint usage of independent P/S ratio estimates at several stations improves the overall performance of the discrimination method. For network discrimination, we use the Mahalanobis distance combining the $P_g/L_g$ values at three stations: USRK, KSRS and MDJ. At frequencies above 4 Hz, the (squared) Mahalanobis distance, $D^2$, between the populations of explosions and aftershocks is larger than 100 with the critical $D^2=16.2$ corresponding to a critical level of significance $\alpha=0.001$ for the chi-square distribution with three degrees of freedom. In the frequency band between 6 and 12 Hz, the aftershocks distance from the average explosion $D^2>21,000$. Statistically, the probability to confuse explosions and aftershocks is negligible.

These discrimination results are related only to the aftershocks of the DPRK tests and cannot be directly extrapolated to the population of tectonic earthquakes in the same area. The diversity of low-magnitude natural and artificial sources within the broader area around the DPRK test site may contain some events much closer to the DPRK explosions than their aftershocks. The population of aftershocks P/S ratios is so tight that it would be instructive to study historical explosions and their aftershocks. We have also to understand better the difference between natural earthquakes and aftershocks. The latter events can be considered as low-magnitude shallow earthquakes with very low stress drop.